\documentclass[english]{cccconf}
\usepackage{epstopdf}
\usepackage{amssymb}
\usepackage{amsmath} 
\usepackage{color}
\usepackage{booktabs} 
\usepackage{hyperref}
\usepackage{url}
\usepackage{cite}
\usepackage{algorithm, float}
\usepackage{algpseudocode}

\newtheorem{theorem}{Theorem}

\newtheorem{definition}{Definition}

\usepackage{caption}
\usepackage{subcaption}

\usepackage{etoolbox}
\usepackage{tikz}
\usetikzlibrary{tikzmark}
\usetikzlibrary{calc}

\usepackage{xcolor}
\errorcontextlines\maxdimen

\newcommand{\ALGtikzmarkcolor}{black}
\newcommand{\ALGtikzmarkextraindent}{4pt}
\newcommand{\ALGtikzmarkverticaloffsetstart}{-.5ex}
\newcommand{\ALGtikzmarkverticaloffsetend}{-.5ex}
\makeatletter
\newcounter{ALG@tikzmark@tempcnta}

\newcommand\ALG@tikzmark@start{%
    \global\let\ALG@tikzmark@last\ALG@tikzmark@starttext%
    \expandafter\edef\csname ALG@tikzmark@\theALG@nested\endcsname{\theALG@tikzmark@tempcnta}%
    \tikzmark{ALG@tikzmark@start@\csname ALG@tikzmark@\theALG@nested\endcsname}%
    \addtocounter{ALG@tikzmark@tempcnta}{1}%
}

\def\ALG@tikzmark@starttext{start}
\newcommand\ALG@tikzmark@end{%
    \ifx\ALG@tikzmark@last\ALG@tikzmark@starttext
    \else
        \tikzmark{ALG@tikzmark@end@\csname ALG@tikzmark@\theALG@nested\endcsname}%
        \tikz[overlay,remember picture] \draw[\ALGtikzmarkcolor] let \p{S}=($(pic cs:ALG@tikzmark@start@\csname ALG@tikzmark@\theALG@nested\endcsname)+(\ALGtikzmarkextraindent,\ALGtikzmarkverticaloffsetstart)$), \p{E}=($(pic cs:ALG@tikzmark@end@\csname ALG@tikzmark@\theALG@nested\endcsname)+(\ALGtikzmarkextraindent,\ALGtikzmarkverticaloffsetend)$) in (\x{S},\y{S})--(\x{S},\y{E});%
    \fi
    \gdef\ALG@tikzmark@last{end}%
}

\apptocmd{\ALG@beginblock}{\ALG@tikzmark@start}{}{\errmessage{failed to patch}}
\pretocmd{\ALG@endblock}{\ALG@tikzmark@end}{}{\errmessage{failed to patch}}
\makeatother

\begin{document}

\title{MPPI-DBaS: Safe Trajectory Optimization  \\ with Adaptive Exploration}

\author{Fanxin Wang\aref{XJTLU},
        Yikun Cheng*\aref{UIUC},
        Chuyuan Tao\aref{UIUC}}

\affiliation[XJTLU]{Department of Department of Mechatronics and Robotics, Xi'an Jiaotong-Liverpool University, P.~R.~China
        \email{Fanxin.Wang@xjtlu.edu.cn}}
\affiliation[UIUC]{Department of Mechanical Science and Engineering, University of Illinois at Urbana-Champaign, USA.
        \email{\{yikun2, chuyuan2\}@illinois.edu}}

\maketitle

\begin{abstract}
In trajectory optimization, Model Predictive Path Integral (MPPI) control is a sampling-based Model Predictive Control (MPC) framework that generates optimal inputs by efficiently simulating numerous trajectories. In practice, however, MPPI often struggles to guarantee safety assurance and balance efficient sampling in open spaces with the need for more extensive exploration under tight constraints. To address this challenge, we incorporate discrete barrier states (DBaS) into MPPI and propose a novel MPPI-DBaS algorithm that ensures system safety and enables adaptive exploration across diverse scenarios. We evaluate our method in simulation experiments where the vehicle navigates through closely placed obstacles. The results demonstrate that the proposed algorithm significantly outperforms standard MPPI, achieving a higher success rate and lower tracking errors.
\end{abstract}

\keywords{Stochastic Optimization, Constraint, MPPI, Barrier State}

\section{Introduction}

With the growing success of robotics applications, the field of robotic control has gained significant attention, leading to the development of more practical strategies for addressing increasingly complex challenges in both simulated and real-world environments. Among these strategies, trajectory optimization is critical, as it aims to determine control inputs that enable a system to achieve specified objectives—such as enhanced energy efficiency and collision avoidance—while adhering to stringent constraints. Model Predictive Control (MPC), a widely adopted optimization-based technique, achieves this by repeatedly solving constrained optimization problems. However, many MPC frameworks assume linear system dynamics and formulate these tasks as quadratic optimization problems. Such simplifications often fail to capture the intricate nonlinearities inherent in robotic systems, resulting in significant model deviations and suboptimal performance.

In contrast, Model Predictive Path Integral (MPPI) control is a sampling-based nonlinear MPC approach that directly accommodates nonlinear system dynamics. By leveraging the parallel computational capabilities of modern CPUs and GPUs, MPPI efficiently samples thousands of trajectories around a nominal control sequence in real time. It then derives an optimal trajectory and corresponding control sequence by computing a weighted average of the sampled trajectories’ costs. A key advantage of MPPI over traditional MPC lies in its ability to handle cost functions without restrictive structural assumptions, allowing for non-quadratic and even discontinuous forms. Nevertheless, most MPC and MPPI implementations represent the robot as a point mass, ignoring its physical geometry, arising the inaccuracy in designing cost functions.
\begin{figure}
  \vspace{4mm}
  \centering
  \scalebox{-1}[1]{\includegraphics[width=0.95\hsize]{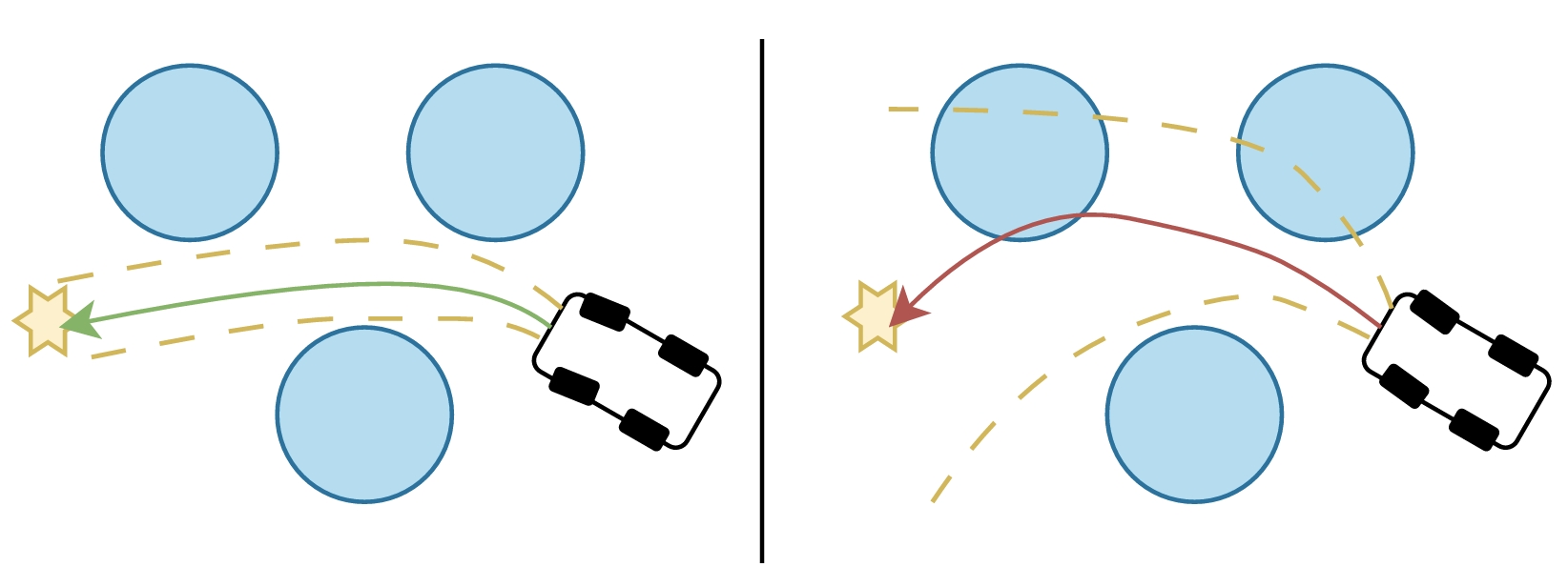}}
  \caption{Demonstration of MPPI trajectory sampling. The slashed yellow curves cover the sampled trajectories. The red curve illustrates a sub-optimal trajectory resulting from extensive, high variance exploration. The green curve denotes a near-optimal, refined trajectory.}
  \label{fig1}
  
  \vspace{-12mm}
\end{figure}

Control Barrier Functions (CBFs) \cite{ames2019control} offer a promising solution to this challenge by representing the robot’s collision boundaries as continuous and differentiable functions that can be integrated into the optimization process as hard constraints. Ensuring system safety via these constraints guarantees that if the system starts in a collision-free set, it will never leave that set under any feasible control sequence derived from the constrained optimization. Such properties have been widely applied in MPC-based methods \cite{zeng2021safety} and learning-based approaches \cite{cheng2023safe}. However, incorporating CBFs introduces additional constraints that increase the complexity of the underlying optimization problem, causing infeasibility and increasing the computational burden.

Barrier states (BaS) \cite{almubarak2023barrier}, a variant formulation of CBFs, address these issues by transforming safety constraints into system stabilization problems rather than embedding them directly into the optimization constraints. This shift simplifies the control scheme design and improves computational efficiency, facilitating more practical real-time implementations for complex robotic systems.

In this work, we propose an MPPI-DBaS framework that ensures robot safety while enabling adaptive exploration through challenging environments, such as navigating narrow yet optimal paths to a destination as illustrated in Figure ~\ref{fig1}. The remainder of this paper is organized as follows: Section~\ref{Sec:prelim} presents the detailed problem formulation and the necessary background on MPPI and DBaS; Section~\ref{Sec:method} describes our proposed algorithm; Section~\ref{Sec:app} illustrates the performance comparison with a standard MPPI approach using a ground vehicle example; and finally, Section~\ref{Sec:con} provides concluding remarks and discusses future research directions.

\section{Preliminaries} \label{Sec:prelim}
In this section, we formalize the optimal control problem and present a detailed overview of the MPPI and DBaS methodologies.

\subsection{MPPI-based Control Problem with Constraints}
We consider the following discrete-time non-linear control system:
\begin{equation}\label{eq:nonmial_dyn}
    x_{k+1} = f(x_k, u_k)
\end{equation}
where $x_k \in \mathbb{R}^n$ represents the $k$th state.
We use $\mathbb{X} = (x_0, x_1, \ldots, x_{N-1}) \in \mathbb{R}^{n \times N}$ to denote the resulting state over the time-horizon $N$, and $\mathbb{U} = (u_0, u_1, \ldots, u_{N-1}) \in \mathbb{R}^{m \times N}$ to denote the control sequence over the time-horizon $N$. 


In trajectory optimization, the objective is to identify a control sequence \(\mathbb{U}\) that transitions the system from an initial state \(x_c\) to a reference state \(x_f\) without incurring collisions. In contrast to gradient-based Model Predictive Control (MPC) approaches, Model Predictive Path Integral (MPPI) does not require the computation of derivatives, making it well-suited for highly nonlinear, non-convex objective functions and complex system dynamics \cite{williams2017model}. The trajectory optimization problem in MPPI can be formulated as:

\begin{equation}\label{eq:mpc_optimization}
    \begin{aligned}
         &\underset{\mathbb{U}}{\text{minimize}} && J= \mathbb{E}[\phi(x_N)+\sum_{k=0}^{N-1}(q(x_k)+\frac{1}{2}v_k^TRv_k)], \\
         &\text{subject to}&&  
    x_{k+1} = f(x_k, v_k), \\
         & && x_0 = x_c, \\
        & && G(u_k) \leq 0, \\
        & && h(x_k) \geq 0, \qquad k\in \{0,1,\cdots,N-1\}
    \end{aligned}
\end{equation}
where $\phi(x_N)$ is the terminal cost, $q(x_k)$ and 
$R$ are the state-dependent running cost and the positive definite control weighting matrix. $v_k = u_k + \delta u_k \in \mathbb{R}^m$ represents the $k$th control input in trajectory sampling. We denote the $\delta u_k \in \mathcal{N}(0, \Sigma_u)$ as the injected disturbance with zero mean and covariance $\Sigma_u$

At each time step, MPPI generates $M$ trajectories in parallel from system dynamics with the injected disturbance  $\delta u_k$ and then evaluates those trajectories based on their expected costs \cite{williams2015model}. The cost-to-go of the state dependent cost of a trajectory is stated as:
\begin{equation}\label{eq:cost-to-go}
\begin{aligned}
    &\Tilde{S}(\tau) = \phi(x_N) + \sum_{k=0}^{N-1} q(x_k)+\gamma u_{k}^T \Sigma_u^{-1} v_k
\end{aligned}
\end{equation}
where $\gamma$ can be interpreted as the control-cost parameter. With $\gamma = 0$, the control sequence \(U\) is push close to zero with high energy conservation. With a large $\gamma$ value, the resulting control sequence \(U\) is the distribution closer to the current planned control law, which keeps \(U\) near the distribution corresponding to \(\hat{U}\). $\Sigma_u$ is again the injected disturbance covariance, which dominates the exploration in parallel trajectory generation.
At $i^th$ step the optimal control sequence $\{u_k\}^{N-1}_{k=i}$ can be estimated as\cite{williams2017model}:
\begin{equation}\label{eq:u_with_rho}
u_k^* = u_k + \frac{\sum_{m=1}^{M}\exp(-(1/\lambda)(\Tilde{S}(\tau_{k,m})-\rho))\delta u_{k,m}}{\sum_{m=1}^{M}\exp(-(1/\lambda)(\Tilde{S}(\tau_{k,m})-\rho))}
\end{equation}
where $\rho = \min_{k} [S_k]$ is the minimum sampled cost in Monte-Carlo approximation, and the parameter $\lambda \in \mathbb{R}^+$
is known as the inverse temperature, which determines the selectivity of the weighted average of the trajectories. The control sequence is then smoothed using a Savitzky-Golay (SG) filter to reduce noise and improve the smoothness of the control inputs. Finally, the first control input $u_0$ in the sequence is applied to the system, while the remaining control sequence of length $N-1$ is used for warm-starting the optimization at the next time-step. 

\subsection{Discrete Barrier States (DBaS)}
By definition, guaranteeing safety is equivalent to rendering the safe set controller invariant over the horizon. 
\begin{definition}\label{def:control-invariant}
    The set $S \subset \mathbb{R}^n$ is said to be controlled invariant with respect to the finite horizon optimal control problem \eqref{eq:mpc_optimization} under feedback policy $U^*(x)$, if $\forall x_0 \in S, x_k \in S  \  \forall \space k \in [0, N] $
\end{definition}
Enforcing controlled invariance can be achieved through various approaches. One such approach is the integration of control barrier functions (CBFs) with trajectory optimization, which effectively addresses both safety and stabilization control objectives in tandem. Nonetheless, CBFs necessitate the precise definition of the relative degree and the establishment of multilevel safe invariant sets. This requirement links safety constraints to control inputs, ensuring safety but also augmenting computational complexity and often resulting in overly cautious system performance. Building upon the concept of using barrier functions to establish CBFs as safety constraints, the method outlined in \cite{khan2020gaussian} incorporates barrier functions directly into the safety-critical system. This concept was later advanced and generalized in \cite{almubarak2022safety}, where barrier states are managed in conjunction with the system's original states. This integration allows safety objectives to be translated into performance objectives within an expanded, higher-dimensional state space. As detailed in \cite{almubarak2023barrier}, the resulting barrier state-embedded model, known as the safety embedded model, achieves asymptotic stability, ensuring safety through the boundedness of the barrier state. Furthermore, discrete barrier states (DBaS) were introduced for discrete-time trajectory optimization, as detailed in \cite{kuperman2023improved} with the application of differential dynamic programming (DDP). Specifically, the barrier function is defined as $B : S \rightarrow \mathbb{R}$ over $h$. A critical aspect of the barrier function $B_h(x_k)$ is its behavior, where $B$ approaches infinity as $x_k$ approaches the boundary of $S$, denoted by $\partial S$.

\begin{definition}\label{def:barrier-state}
    A scalar output function $\boldsymbol{B}:\mathcal{S}\to\mathbb{R}$ is a Barrier Function (BF), if it is smooth and strongly convex on the interior of the set of feasible solution $\mathcal{S}$ and blows up as its input approaches the boundary of the set $\partial\mathcal{S}$. Mathematically, for a sequence of inputs $\{\eta_1,\eta_2,...,\eta_k\}$,
    \begin{equation*}
        \eta_k \in \mathcal{S}, \tilde{\eta} \equiv \lim _{k \rightarrow \infty} \eta_k \in \partial \mathcal{S} \Rightarrow \mathbf{B}\left(\eta_k\right) \rightarrow \infty, k \rightarrow \infty\\
    \end{equation*}
    with
    \begin{equation*}
        \inf _{\eta \in \mathbb{R}^{n+}} \mathbf{B}(\eta) \geq 0.
    \end{equation*}
\end{definition}

With this, We define the barrier function $\beta(x_k):=\boldsymbol{B}\circ h(x_k)$. As a consequence, by definition, ensuring the barrier function $\beta$ is bounded implies the satisfaction of safety condition $h(x_k) \geq 0$. With the barrier state defined above, we define the safety embedded system with $\hat{x} = \begin{bmatrix} x \\ \beta \end{bmatrix}$:
\begin{equation}\label{eq:safety_embedded_dyn}
    \hat{x}_{k+1} = \hat{f}(\hat{x}_k, u_k)
\end{equation}

\begin{theorem}\label{thm1}
    For the safety embedded system \eqref{eq:safety_embedded_dyn}, given $x(0)\in\mathcal{S}$, a continuous feedback controller $u=K(x)$ is safe, i.e. satisfies the safety condition $h(x_k) \geq 0$ and the safe set $\mathcal{S}$ is controlled invariant, if and only if $\beta(x_k)<\infty,\forall k\in \{0,1,\cdots,N-1\} $.
\end{theorem}

To ensure the stability of the safety embedded system, the barrier state equation was proposed by \cite{almubarak2023barrier} as 
\begin{equation}\label{eq:barrier-state}
\beta(x_{k+1})=\mathbf{B}\circ h \circ  f(x_k, u_k)-\gamma(\mathbf{B} \circ h(x_d)-\beta(x_k)),
\end{equation}
where $\gamma \in (0,1)$  is a tunable parameter. $x_d$ is the desired equilibrium point. It is worth noting that the DBaS equation is a function with respect to $\beta_k$, with the $\gamma$ item embedded, it ensures the converging reduction of the state estimation error and analogous to the observer gain in the dynamic compensation. 
 
For multiple constraints, one single barrier state could be formulated to ensure the safety from multiple constraints \cite{almubarak2023barrier}:
 \begin{equation}\label{eq:multi-constraints}
w(x_{k+1})=\sum_{i=1}^{n_\beta} \mathbf{B} \circ h^{(i)} \circ  f(x_k, u_k)-\gamma(\beta(x_d)-\beta(x_k)),
\end{equation}
where the barrier state vector $\beta(x_k) =\sum_{i=1}^{n_\beta} \mathbf{B} \circ h^{(i)} (x_k)$, $ \space  \beta \in \mathcal{B} \subset \mathbb{R}^{n_\beta}$, where $n_\beta$ is the dimensionality of the barrier state vector. $w$ is now the fused barrier state with all constraints embedded and it would be appended to the dynamical model resulting in the safety embedded system:

\begin{equation}\label{eq:augment-BaS-dyn}
\hat{x}_{k+1} = \hat{f}(\hat{x}_k, u_k)
\end{equation}
where $\hat{x} = \begin{bmatrix} x \\ w \end{bmatrix}$ and $\hat{f} = \begin{bmatrix} f \\ f_w \end{bmatrix}$.

By incorporating the barrier state as shown in Equation \eqref{eq:augment-BaS-dyn}, the dynamic system now inherently incorporates safety constraint information into its state representation. This integration transforms the conventional approach, which typically involves separate optimal and safety control algorithms competing for control bandwidth, into a cohesive framework that eliminates the need for such contention.

\section{MPPI-DBaS with adaptive exploration} \label{Sec:method}

\begin{figure*}[!ht]
  \centering
  \includegraphics[width=0.7\hsize]{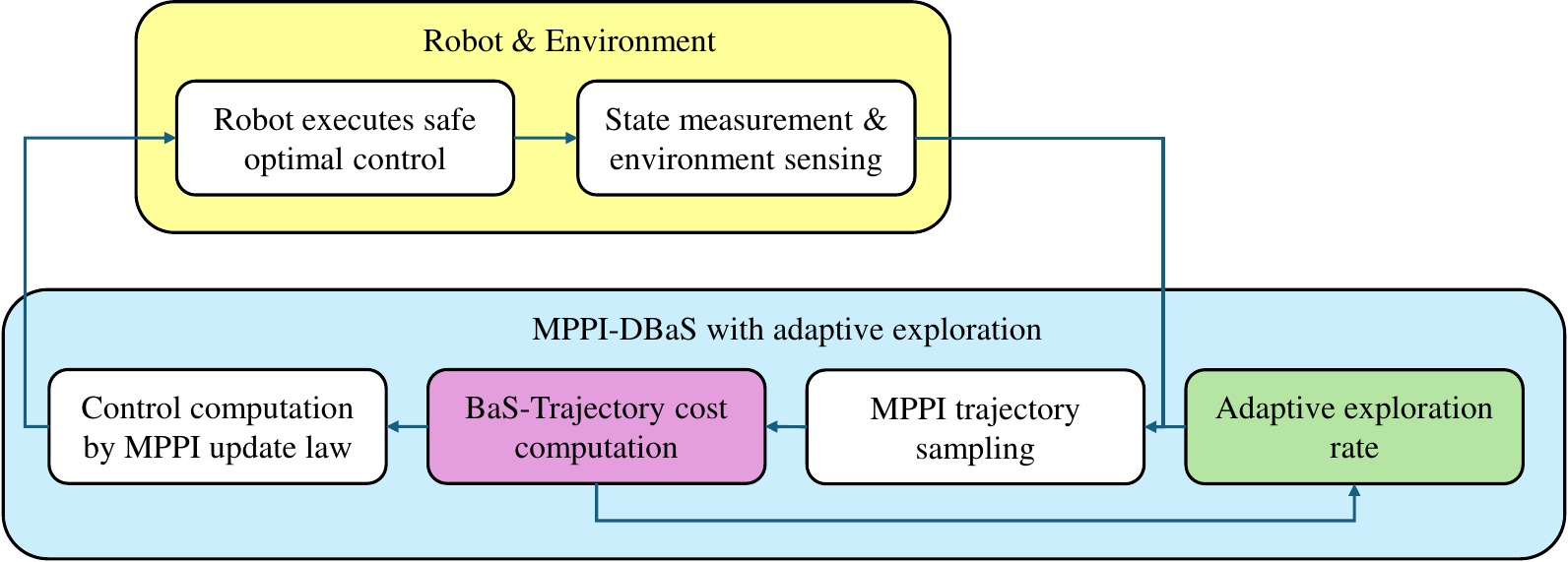}
  \caption{Proposed MPPI-DBaS control scheme.}
  \label{fig2}
  \vspace{-5mm}
\end{figure*}

Incorporating safety constraints into a standard MPPI controller presents a significant challenge. It normally requires a long planning horizon to ensure the system safety well into the future, to avoid the sudden transition from safe to unsafe states. As the MPPI is a sampling-based optimization algorithm, the requirement for a lengthy planning horizon significantly increases the computational load in evaluation, especially when the MPPI controller is also tasked with extended trajectories to identify near-optimal actions.

To alleviate the computational cost of deploying a standard MPPI with constraints, we have developed the MPPI-DBaS controller, featuring two key enhancements to the control strategy.

Firstly, to enable the controller to maintain safety with a shorter planning horizon, we integrate barrier states into the system's dynamics, thereby addressing the safety concerns as part of the states under control. Additionally, we have incorporated the cost of these barrier states into the running cost, replacing the standard collision impulse-indicator cost typically used in MPPI \cite{yin2023shield}.

Secondly, we introduce an adaptive exploration mechanism of sampling trajectories. Given that the barrier state ensures the safety of each sampled trajectory, a high running cost would evaluate a closer proximity to obstacles and a higher risk of collision. In return, this guide our exploration strategy. To avoid the local optima commonly met without a fine-tuned safety cost, we would employ a higher exploration rate that encourages more diverse trajectory sampling, yielding roughly optimal paths \cite{yin2022trajectory}. Subsequent action steps, informed by this initial broad search, can then commence with a lower exploration rate, leading to a more precise near-optimal solutions with safety assurances.

In this section, we will introduce the detailed algorithm of MPPI-DBaS with its enhanced safety guarantees. The block diagram could be referred to Figure ~\ref{fig2}.

\subsection{DBaS augmentation with the running cost}
In the standard MPPI, constraints are assessed using weighted indicator terms. These costs impose an impulse-like penalty upon constraint violation, serving as a hybrid approach between hard and soft constraints. However, this method does not provide any risk information until the constraints are violated, posing challenges in trajectory optimization. Specifically, until an actual collision is predicted to occur, all trajectories are evaluated solely based on state-dependent costs. This approach does not ensure safety in scenarios characterized by narrow passages or high speeds. This could be referred to Figure ~\ref{fig3a} and Figure ~\ref{fig3b} in the result section. Thus, we propose the following DBaS augmentation in the running cost.

Starting at each loop, the adaptive exploration rate (which would be introduced in the next subsection) would determine the diversity in trajectory sampling. $\delta u_k$ as well as the corresponding trajectories would be sampled at the desired exploration rate with the DBaS augmented system dynmacis:

\begin{equation}\label{eq:MPPI_BaS}
    \begin{aligned}
        &\hat{x}_{k+1} = \hat{f}(\hat{x}_k, v_k) \\
        &v_k = u_k + \delta u_k
    \end{aligned}
\end{equation}
where $\hat{x}_k$ is the introduced new system state embedded with DBaS. 
Since the stochastic sampling of candidate trajectories and subsequent cost evaluations remain unchanged, the convergence behavior of MPPI is preserved\cite{williams2018information}, as DBaS does not compromise the asymptotic convergence properties of the MPPI algorithm. By seamlessly integrating safety considerations into the cost structure, DBaS ensures strict constraint satisfaction without undermining the stability, robustness, or convergence guarantees inherent to the underlying optimization framework.

Furthermore, DBaS offers a distinct advantage in controlling high relative-degree systems, where constraints at the output level are influenced only through multiple derivatives of the control input. Traditional approaches often require defining barrier functions at higher derivatives and incorporating carefully chosen exponential terms to achieve exponential stability of constraint enforcement.
In contrast, DBaS leverages costate dynamics to naturally encapsulate the interdependencies between states and constraints. Theorem~\ref{thm1} provides a necessary and sufficient condition to guarantee safety in high relative-degree systems, streamlining the synthesis process and reducing implementation complexity. As a result, DBaS not only simplifies the enforcement of safety constraints in complex, high-order dynamical systems but also enhances the applicability by avoiding scalability issues.

By sampling the barrier state at each time step, we can ensure that safety constraints are satisfied throughout the trajectory, while avoiding the computational complexity of solving the barrier state equation at each iteration. This approach significantly improves the efficiency of the controller and guarantees that the system stays within safe bounds during real-time operation in dynamic environments.

$M$ number of samples are generated under this new dynamics, and we would calculate the new cost-to-go of the state dependent cost of a trajectory:
\begin{equation}\label{eq:cost-to-go_bas}
    \Tilde{S}(\tau) = C_B(\hat{X})+\phi(\hat{x}_N) + \sum_{k=0}^{N-1} q({x}_k)+\gamma u_{k}^T \Sigma_u^{-1} v_k
\end{equation}
where the term $C_B(\hat{X}) = \sum_{k=0}^{N} R_B \space w({x}_k)$ denotes the barrier state cost along the trajectory. $R_B$ is the cost weight on the barrier state, and we do not need the square of the barrier state to ensure the positive definiteness on the term since the barrier state itself is always positive. And this cost-to-go function also implies penalizing collision in the sampled trajectory even though only the first control instance is applied on the system \eqref{eq:MPPI_BaS}. 

With equation \eqref{eq:u_with_rho}, we could create a sequence of "near-optimal" controls $\mathbb{U_k^*} = (u_0^*, u_1^*, \ldots, u^*_{N-1}) \in \mathbb{R}^{m \times N}$ that would be applied to the
system \eqref{eq:MPPI_BaS} optimizing the cost $J= \mathbb{E}[\Tilde{S}(\tau)]$.

\subsection{Adaptive trajectory sampling}
In the standard MPPI, $M$ trajectories are sampled with the same covariance to obtain the "near-optimal" controls. This guarantees great computation speed without finding the explicit gradients as required in the optimization problem. However, in tightly-constrained system-environment scenario, applying the same exploration rate determined by the same $\Sigma_u$ would lead to two possible issues:

On one hand, setting $\Sigma_u$ to a relatively small value makes it easier for MPPI to evaluate and decide a fine trajectory among the samples in a loosely-constrained task, as most trajectories are in close proximity. However, this could lead to entrapment in a local optimum, given the limited exploration through closely clustered samples in a tightly-constrained system-environment scenario. For further illustration, see Figure ~\ref{fig3c} in the results section.

On the other hand, setting $\Sigma_u$ to a relatively large value increases the likelihood of finding a path through a narrow gap in a tightly-constrained system-environment scenario, as the $M$ sampled trajectories explore a greater portion of the hyper space of the controller. However, this optimization approach may under-perform in path following in open spaces, as the trajectories are widely dispersed, leading MPPI to tend to identify a coarse trajectory rather than a precise one with accurate tracking. For more details, refer to Figure ~\ref{fig3d} in the results section.

Thus, we propose an adaptive trajectory sampling framework that leverages the DBaS property to ensure constraint satisfaction. As detailed in Section 2.2, by embedding DBaS within the nominal state, we guarantee that constraints will not be violated. Consequently, the also grants us the flexibility to increase the value of $\Sigma_u$ when dealing with tight constraints. When approaching the forbidden area, the value of $h(x_k)$ would decrease, and according to Definition~\ref{def:barrier-state}, the $B\circ h(x_k)$ would increase to a very large value resulting in the large value of $w(\hat{x}_k)$. We design the following exploration rate, $S_e$, such that the injected disturbance $\delta \hat{u}_k$ follows a normal distribution $\mathcal{N}(0, S_e\Sigma_u)$, ensuring the disturbance has zero mean and covariance $S_e\Sigma_u$.
\begin{equation}\label{eq:scaling factor}
    S_e = \mu \space log(e+ C_B(\hat{X}^*))
\end{equation}
where $\mu \in (0,1)$ is the coarseness factor, determining  the accuracy of the trajectory while tracking in free space. As the barrier state cost along the trajectory $ C_B(\hat{X}^*)>0$, we have $S_e>\mu, \forall X^*\in \mathbb{R}^{n \times N}$. The logarithmic function is applied to $S_e$ to ensure a non-aggressive exploration behavior, as the cost increases rapidly when approaching obstacles or limits in a tightly-constrained scenario.

$S_e$ would be updated after every optimization step, applying the current $\mathbb{U}^*$ for sampling the 'near-optimal' trajectory and evaluating barrier cost $ C_B(\hat{X}^*)$ for that trajectory.

\subsection{MPPI-DBaS with adaptive exploration algorithm}
Same as suggested in the Figure~\ref{fig2}, we propose the detailed MPPI-DBaS with adaptive exploration algorithm as the following:

\begin{algorithm}[!htbp]
\caption{Sampling-Based MPC.}
\begin{algorithmic}[1]
\Require $f$: Transition Model; \\
$M$: Number of samples; \\
$T$: Number of time-steps; \\
$(u_0, u_1, \ldots, u_{T-1})$: Initial control sequence; \\
$\mu, \Sigma_{u}, \phi, q, \gamma, R_B$: Cost functions/parameters; \\
SGF: Savitsky-Golay convolutional filter;
\While{task not completed}
    \State $x_0 \gets \text{GetStateEstimate}()$
    \For{$k \gets 0$ to $M-1$}
        \State $x \gets x_0$
        \State Sample $ (\delta u_0^k, \ldots, \delta u_{T-1}^k)$, $\delta u_t^k \sim \mathcal{N}(0, S_e\Sigma_u)$
        \For{$t \gets 1$ to $T$}
            \State $v = u+\delta u$
            \State $x \gets f(x, v)$
            \State $w \gets \sum\mathbf{B} \circ h^{(i)} \circ  f(x_t, v_t)-\gamma(\beta(x_d)-\beta(x_t))$
            \State $\hat{x} = \begin{bmatrix} x \\ w \end{bmatrix}$
            \State $\Tilde{S}_k += q(\hat{x})+\gamma u_{}^T S_e^{-1}\Sigma_u^{-1} v+ R_B \space w({x}_k)$ 
        \EndFor
        \State $\Tilde S_k += \phi(x)$
    \EndFor
    \For{$t \gets 0$ to $T-1$}
        \State $U^* \gets U + \text{SGF}  \frac{\sum_{m=1}^{M}\exp(-(1/\lambda)(\Tilde{S}_k-\rho))\delta u_{k,m}}{\sum_{m=1}^{M}\exp(-(1/\lambda)(\Tilde{S}_k-\rho))}$
    \EndFor
        \State SendToActuators($u_0$)
    \For{$t \gets 1$ to $T$}
        \State $\hat{x}= \hat{f}(\hat{x}, u*) $
        \State $C_B(\hat{X}^*)+=R_B \space w({x}_k)$
    \EndFor
    \State $S_e = \mu \space log(e+ C_B(\hat{X}^*))$
    \For{$t \gets 1$ to $T-1$}
            \State $u_{t-1} \gets u_t$
        \State $u_{T-1} \gets \text{Initialize}(u_{T-1})$
    \EndFor
\EndWhile
\end{algorithmic}
\end{algorithm}

\section{Applications} \label{Sec:app}

\begin{figure}[!t]
\vspace{-8mm}
     \centering
     \begin{subfigure}{0.235\textwidth}
         \centering
         \includegraphics[width=\textwidth]{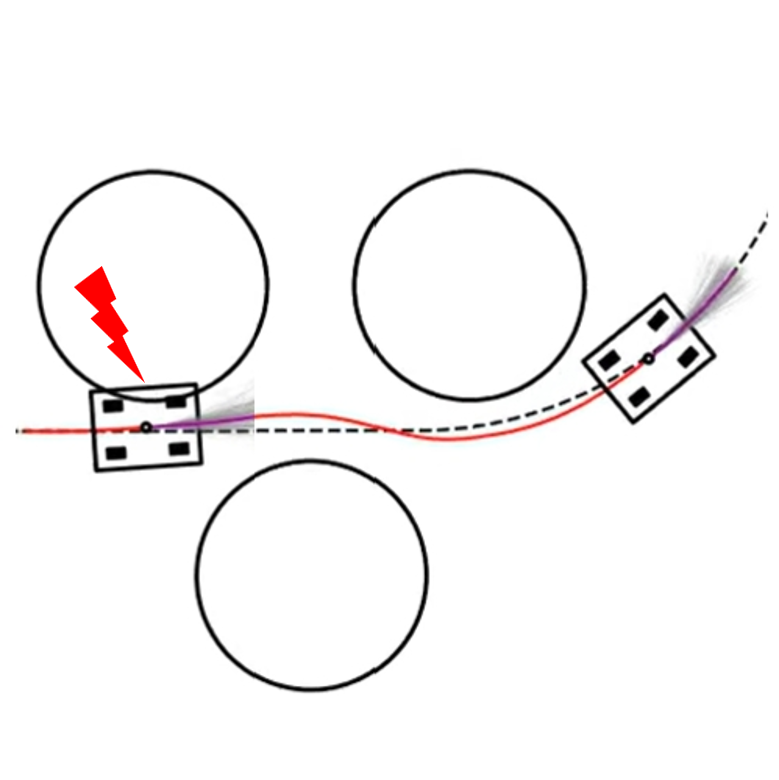}
         \caption{Collision at first obstacle.}
         \label{fig3a}
     \end{subfigure}
     \hfill
     \begin{subfigure}{0.235\textwidth}
         \centering
         \includegraphics[width=\textwidth]{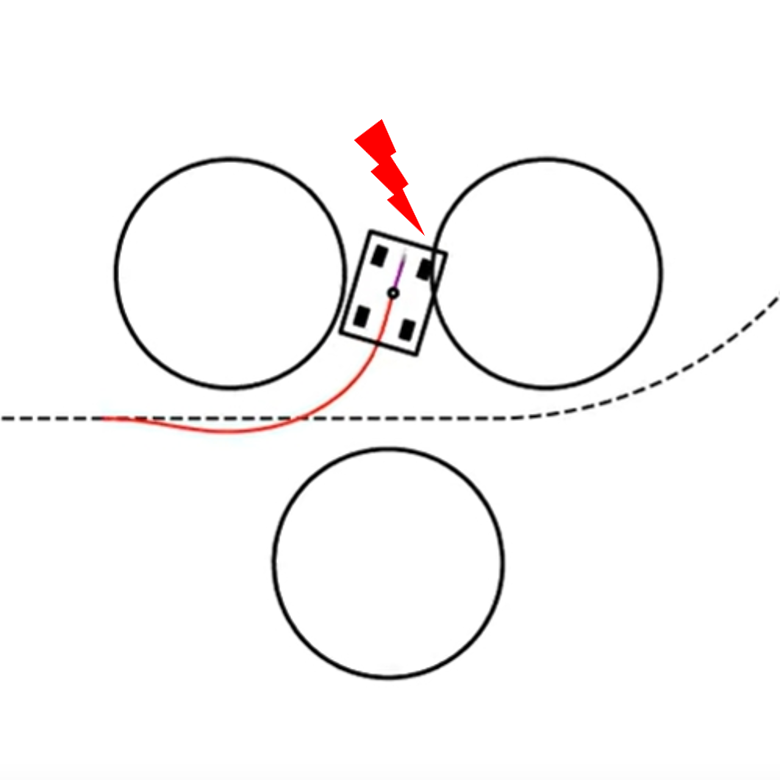}
         \caption{Collision at third obstacle.}
         \label{fig3b}
     \end{subfigure}
     \hfill
     \begin{subfigure}{0.235\textwidth}
         \centering
         \includegraphics[width=\textwidth]{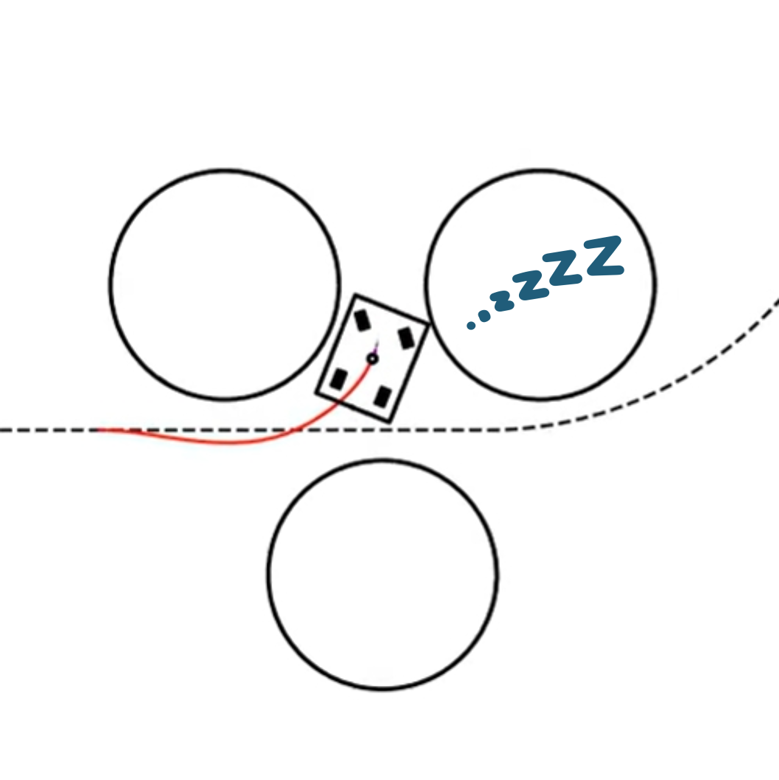}
         \caption{Entrapped in local optima.}
         \label{fig3c}
     \end{subfigure}
     \hfill
     \begin{subfigure}{0.235\textwidth}
         \centering
         \includegraphics[width=\textwidth]{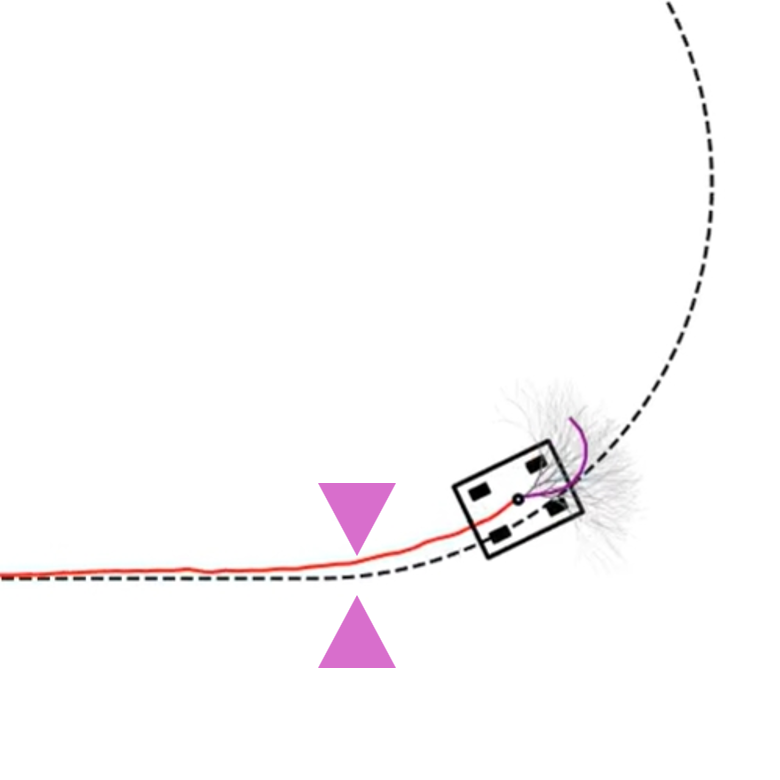}
         \caption{Deviation from reference.}
         \label{fig3d}
     \end{subfigure}
         \caption{MPPI fail cases. The sampled trajectories are depicted as grey curves, with the near-optimal trajectories highlighted in purple. The red curves indicate the actual trajectory that has been executed.}
        \label{fig3}
\vspace{-8mm}
\end{figure}
In this section, we present simulation results comparing the proposed MPPI-DBaS controller with the standard MPPI on a path-tracking vehicle. (A video can be found: \url{https://youtu.be/lwwFcYQTql4}) The vehicle is modeled as the following discrete-time nonlinear Ackermann steering model:
\begin{equation*}\label{veh_dyn}
    \begin{bmatrix}
x_{k+1}\\ 
y_{k+1}\\ 
\theta_{k+1}\\
v_{k+1}
\end{bmatrix} = \begin{bmatrix}
x_{k} + v_k\cos\theta_k\Delta t\\ 
y_{k} + v_k\sin \theta_k\Delta t\\ 
\theta_{k} +  \frac{v_k\tan \phi}{L}\Delta t\\
v_k+a\Delta t
\end{bmatrix},
\end{equation*}
where $x_{k}, y_{k}, \theta_{k}, v_k$ denote the Cartesian positions, yaw and linear velocity of the vehicle at $k$th timestamp. $\phi, a$ denote the control signal of steering angle and acceleration. The planning optimization problem \eqref{eq:mpc_optimization} is designed such that the robot must navigate while avoiding circular obstacles and tracking the reference path. We propose a specific vehicle model for this purpose, where collision detection is facilitated by defining eight shape points on the vehicle's geometric model—a rectangle with a length of 4 meters and a width of 3 meters. These shape points are positioned at the corners and midpoints along each side of the rectangle. The obstacle function is defined as:
\begin{equation*}
    \begin{aligned}
   h(x_i) =\| x_i - x_c^j \|_2^2 - r_c^j{^2} \geq 0, \quad \text{for } &i \in \{ 1, \dots \, 8\}, \\
   &j\in \{1, \dots \}
   \end{aligned}
\end{equation*}
where $x_i$ denotes the shape points and $x_c^i$ is the center of $i$th circle obstacle with radius of $r_c$.

The vehicle is tracking a path that begins with a straight line followed by a semi-circle at a reference speed of 5 m/s. Multiple obstacles with narrow gaps are placed along the path to evaluate the capabilities of both the standard MPPI and the proposed MPPI-DBaS controller. Sampling covariance matrix $\Sigma_u$ are both as $\begin{bmatrix} 0.075& 0 \\ 0& 2 \end{bmatrix}$, $\gamma$ are set to be 2. Coarseness factor $\mu$ is set to be 0.4 in MPPI-DBaS.

\begin{figure}[!b]
     \centering
     \begin{subfigure}{0.235\textwidth}
         \centering
         \includegraphics[width=\textwidth]{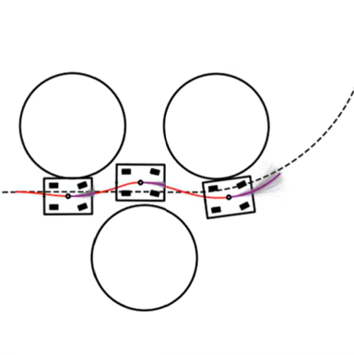}
         \caption{Safe collision avoidance.}
         \label{fig4a}
     \end{subfigure}
     \hfill
     \begin{subfigure}{0.235\textwidth}
         \centering
         \includegraphics[width=\textwidth]{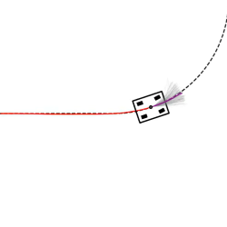}
         \caption{Accurate \& adaptive track.}
         \label{fig4b}
     \end{subfigure}
         \caption{MPPI-DBaS successful tracking. The sampled trajectories are depicted as grey curves, with the near-optimal trajectories highlighted in purple. The red curves indicate the actual trajectory that has been executed.}
        \label{fig4}
\end{figure}

In the standard MPPI, safety is assessed using an indicator function, which does not evaluate risk until the collision in the prediction period. As illustrated in Figures ~\ref{fig3a} and~\ref{fig3b}, collisions can occur during sampling and exploration. Furthermore, with a fixed exploration rate determined by the sampling covariance matrix, the vehicle may become trapped in local optima, as shown in Figure~\ref{fig3c}. Conversely, if $\Sigma_u$ is set to a larger value, the vehicle may underperform, resulting in significant positional errors even in the absence of obstacles as depicted in Figure~\ref{fig3d}.

On the other hand, the MPPI-DBaS proposed in this paper incorporates a barrier state into the system dynamics, as shown in \eqref{eq:MPPI_BaS}. This approach is capable of handling tight-constrained scenarios with continuous risk awareness as shown in Figure~\ref{fig4a}. Furthermore, with an adaptive exploration rate applied to the covariance matrix, the sampled trajectories will diverge broadly when approaching obstacles, ensuring comprehensive coverage, and concentrate in open areas, maintaining efficiency as shown in Figure~\ref{fig4b}.

Table~\ref{tab:table1} summarizes the simulation results which carried out 20 times for each controller. It demonstrates that the MPPI-DBaS controller achieves full success rate at relatively high velocities while maintaining low position errors in general, while standard MPPI faces challenges in tightly-constrained scenarios.

\begin{table}[!h]
    \caption{Comparison between MPPI-DBaS and baseline MPPI. Simulation are carried out 20times in both cases.}
    \begin{center}
        \begin{tabular}{ccc}
        \toprule[1pt]
        Controllers & MPPI-DBaS & Standard MPPI\\ \midrule
        Success times & 20 & 1 \\
        Fail(stop) times & 0 & 11 \\ 
        Fail(collision) times & 0 & 8 \\
        Avg vel(m/s) & 4.13 & 3.21 \\
        Avg pos error(m) & 1.21 & 1.87 \\
        \bottomrule[1pt]
        \end{tabular}   
    \end{center}
    \label{tab:table1}
\end{table}

\begin{figure}[!h]
     \centering
     \begin{subfigure}{0.3\textwidth}
         \centering
         \includegraphics[width=\textwidth]{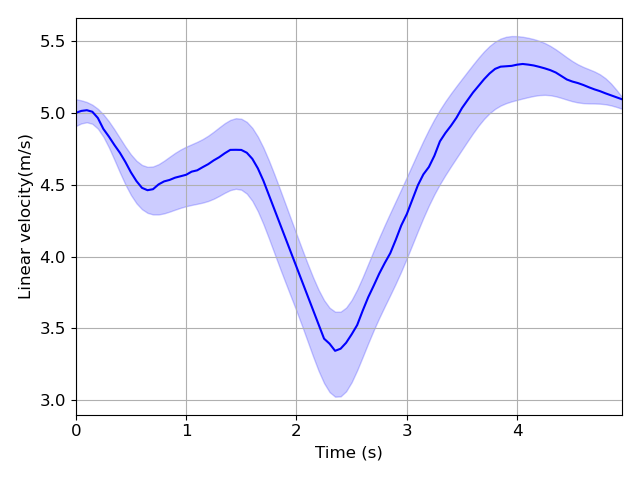}
         \caption{Simulation results of velocity.}
         \label{fig5a}
     \end{subfigure}
     \hfill
     \begin{subfigure}{0.3\textwidth}
         \centering
         \includegraphics[width=\textwidth]{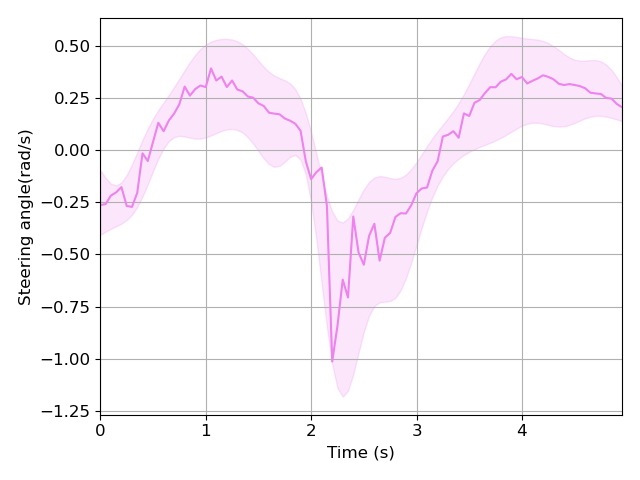}
         \caption{Simulation results of steering angle.}
         \label{fig5b}
     \end{subfigure}
     \hfill
     \begin{subfigure}{0.3\textwidth}
         \centering
         \includegraphics[width=\textwidth]{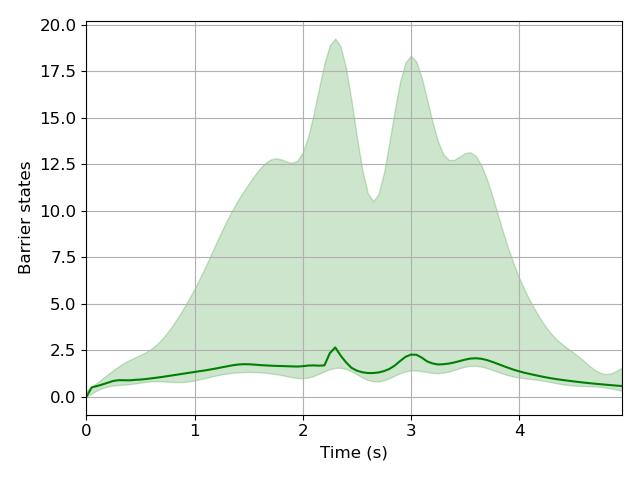}
         \caption{Simulation results of Barrier states.}
         \label{fig5c}
     \end{subfigure}
         \caption{State and control result from 20 simulations. The mean is traced with solid line in each figure, with the variation is depicted by a shaded area.}
        \label{fig5}
\end{figure}

As the proposed MPPI-DBaS is a sampling-based optimization scheme, its performance may exhibit slight variations across simulations, even under identical conditions (particularly when employing a small number of \( M \) sample trajectories). Figure~\ref{fig5} illustrates these performance variances, focusing on linear velocity, steering angle, and the barrier state. The barrier state shows the most significant variation, as even minor approaches to obstacle due to variation in close distances can result in substantial changes in its value.

\section{Conclusion} \label{Sec:con}
In this paper, we introduce a novel MPPI-DBaS controller that integrates a barrier state into the dynamic system, thereby ensuring hard-safety capabilities, surpassing the standard indicator function used in standard MPPI. Additionally, we have developed an adaptive exploration mechanism for sampling trajectories, which facilitates more extensive and varied sampling when approaching obstacles. Our simulations and experimental results demonstrate that the proposed algorithm significantly enhances the success rate with lower tracking error, compared to standard MPPI algorithms, particularly in scenarios with strict constraints.

 In the future, we aim to refine the MPPI-DBaS controller by developing safety barriers in more comprehensive forms and applying these enhanced algorithms to more complex control scenarios, including multi-agent planning \cite{zhang2024gcbf+}. Additionally, we plan to leverage GPU-based parallel computing \cite{tao2022path} to sample a greater number of trajectories at a higher computational speed, which will facilitate more exhaustive exploration within our sampling process.

\bibliographystyle{ieeetr}
\bibliography{Feifei}

\end{document}